\documentclass[a4,aps,amsmath,floatfix,nofootinbib,10pt]{revtex4-2} 
\usepackage{graphicx} \usepackage{color} \usepackage{enumerate} 
\usepackage{amssymb} \usepackage[T1]{fontenc}
\usepackage{ae}
\newcommand{\be}{\begin{equation}} \newcommand{\ee}{\end{equation}} 
\newcommand{\ba}{\begin{array}} \newcommand{\ea}{\end{array}} 
\newcommand{\bea}{\begin{eqnarray}} \newcommand{\eea}{\end{eqnarray}} 
\newcommand{\bdm}{\begin{displaymath}} 
\newcommand{\edm}{\end{displaymath}}

\usepackage[colorlinks]{hyperref} 
\usepackage{float}

\begin{document}

\title{Monte Carlo Simulation of Anisotropic Ising Model Using Metropolis and Wolff Algorithm}

\author{Basit Iqbal$^{1}$, Kingshuk Sarkar$^{* 1}$}

\affiliation{$^{1}$Department of Physics, School of Advanced Sciences, VIT-AP University, Amravati, Andhra Pradesh – 522237, India.\\
	*kingshuk.sarkar@vitap.ac.in}

\begin{abstract}

We employ Monte Carlo techniques, utilizing the Metropolis and Wolff algorithms, to investigate phase behavior and phase transitions in anisotropic Ising models. Our study encompasses the thermodynamic properties, evaluating energy, magnetization, specific heat, magnetic susceptibility, magnetic entropy, and the Binder cumulant. Additionally, we examine the impact of external fields on these thermodynamic quantities at different externally applied field values. We accurately determine the critical temperature for various model scenarios by analyzing the Binder cumulant. Our investigations also include an analysis of the hysteresis loop for the model for different an-isotropic cases. In particular, our study presents the magnetocaloric effect, which is the change in temperature of magnetic material when exposed to a changing magnetic field, in the different anisotropic cases of the Ising model.

\end{abstract}

\maketitle

\section{introduction}
\label{S_1}
\begin{figure}[ht] 
	\centering
	\includegraphics[width=8cm,height=7cm]{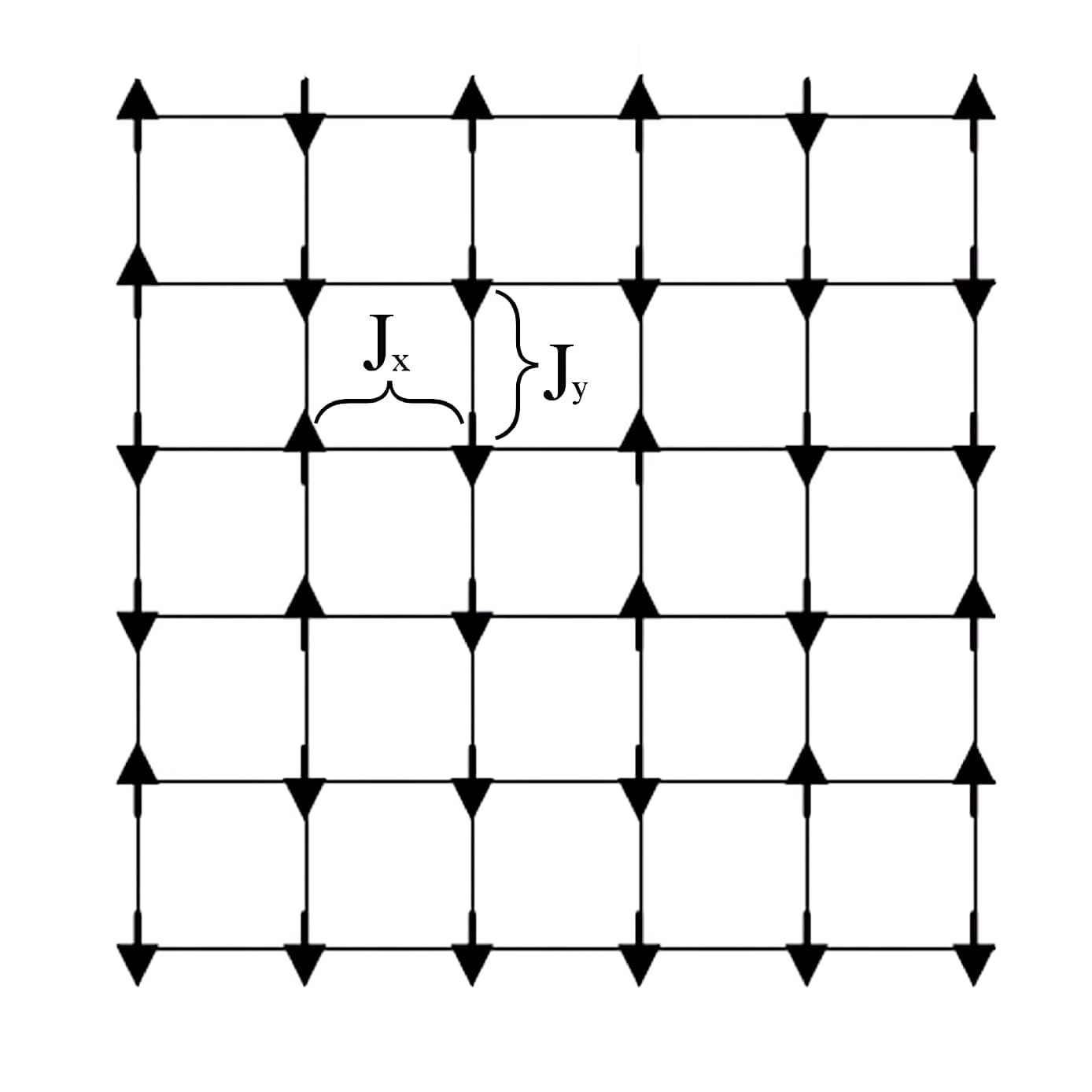}.\\
	\caption{Schematic diagram of 2D Ising model with $J_x$ and $J_y$ as nearest neighbor interactions along x-axis and y-axis respectively.}
	\label{Figure_1}
\end{figure}

The main aim of condensed matter physics is the study of the collective properties of matter composed of many constituents and the interactions between them~\cite{chaikin_lubensky}. Distinguishing different phases of a system and locating the critical points in the parameter space is key in condensed matter and statistical physics. The study of magnetic materials is a significant topic of interest in both fields. This magnetism arises from the spins of electrons within the material, and ferromagnetic materials particularly are important to study. Ferromagnetic materials have domains in which spins are aligned in a specific direction. When an external magnetic field is applied to these materials, the domains align with the applied field, magnetizing the material. As the temperature increases, the total magnetization decreases. At a specific temperature, known as the Curie temperature or transition temperature, the magnetizations become disordered, and the spins point in random directions, causing the material to lose its magnetic properties--the transition from the ferromagnetic to the paramagnetic state. This study focuses on explaining the thermal and magnetic behavior of magnetic materials.\\
In 1925, a German physicist, Ernst Ising, with the advice of his PhD advisor Wilhelm Lenz, discussed a model known as the Ising model in his PhD thesis~\cite{ising_1925}. The Ising model was later solved exactly by Onsager for two dimensions~\cite{onsager_1944}. This model is considered the paradigmatic model for the study of equilibrium phase transitions in statistical and condensed matter physics, as well as in various other fields such as chemistry, biology, statistics, and finance~\cite{stauffer_2007,thompson_1979,chernodub_2010,kovacs_2009,bulled_2022,yilmaz_2005,kofinger_dellago_2010,azbel_1979,jon_1970,gibbs_1948,sneppen_zocchi_2005,bakk_hoye_2003}.\\
The Ising model is one of the simplest and fundamental models for studying magnetic materials. The model has been extensively studied due to its phase transition properties, and numerous resources are available, such as~\cite{pathria_beale_2011}. An overview of the model is as follows. Our study considers a square lattice of size $L$, having $N(=\mathrm{L \times L})$ lattice sites. The symbol $s_{i}$ (or $\sigma_{i}$) is used to represent the magnetic moment (or spin) of the electron/atom/molecule at the lattice site $i$. The spin is considered fixed at the site and is assumed to take only binary values $\{1,-1\}$ corresponding to the two possible spin states, up and down, along a fixed axis. See Fig.\ref{Figure_1}. The choice of spins with values of $\pm 1$ is foundational to the model and is motivated by several reasons. This binary nature simplifies the mathematical formulation and analysis of the model, making it more tractable for theoretical studies and computational simulations. The model was originally developed to describe ferromagnetic materials, where atomic magnetic moments (spins) tend to align either parallel or anti-parallel. The Ising model with $\pm 1$ spins has been instrumental in understanding critical phenomena and phase transitions, particularly second-order phase transitions. These sites ($i$ and $j$) interact through their fields with interaction strengths $\mathrm{J}_{\mathrm{x}}$ or $\mathrm{J}_{\mathrm{y}}$ along x-axis and y-axis respectively. Furthermore, the interactions are taken between nearest neighboring sites, denoted as $\langle i, j\rangle$ under the summation. The Hamiltonian for the Ising model is represented as:
\begin{equation}
	\mathrm{H}(\mathrm{s})=-\mathrm{J}_{\mathrm{x}} \sum_{\langle i,j \rangle_{x}} s_{i} s_{j}-\mathrm{J}_{\mathrm{y}} \sum_{\langle i, j\rangle_{y}} s_{i} s_{j}-h_j \sum_{j} s_{j} 
	\label{eq_1}	
\end{equation}
Here, $h_j$ is the external magnetic field applied. Spins oriented along the magnetic (external) field will have lower energy, and those oriented opposite to the field will have higher energy. For $\mathrm{J}_{x}, \mathrm{J}_{y}>0$, neighboring spins tend to align in a single direction and material shows ferromagnetic behavior. Conversely, for $\mathrm{J}_{x}, \mathrm{J}_{y}<0$, the material becomes antiferromagnetic. When $\mathrm{J}=0$, the spins do not interact with each other. In these simulations, $J$ is set as the unit of energy and temperature. The changes in the value of J is explained ahead for different anisotropic scenarios.\\
By the term anisotropic Ising model, we mean that  $\frac{J_y}{J_x} \ne 1.0$. This is the model we are interested in our simulations. In the isotropic Ising model, the values of $\frac{J_y}{J_x} = 1.0$. In our study of the anisotropic Ising model, the value of $\frac{J_y}{J_x}$ has been changed. One of the methods for studying the phase transition of the model is the Monte Carlo (MC) technique. Metropolis algorithm has been implemented successfully on a large variety of condensed matter systems like cuprates~\cite{banerjee_2018,sarkar_2016,sarkar_2017}. On the one hand, following the tradition of using the Monte Carlo method, we employed the Metropolis-Hastings algorithm (discussed in section \ref{MHA} ahead) to study phases and phase transitions of the 2D Ising model. However, this approach leads to the well-known problem of critical slowing down near the critical point. This problem has been extensively discussed in the literature~\cite{landau_paez_2007}. While we will not delve deeply into the details of this issue, we aim to give the reader a sense of its complexity in the next few paragraphs.\\
On the other hand, we used the Wolff algorithm~\cite{wolff_1989} for the same study of phases and phase transitions. The Wolff algorithm helps us bypass critical slowing down by employing a method called cluster updating, which involves growing a cluster randomly and flipping the cluster, giving the speed-up over the Metropolis-Hastings' algorithm which flips a single spin randomly. The working of the Wolf algorithm has been discussed in section \ref{WA} ahead.\\
2D materials in general and the ones with magnetic properties especially, after the discovery of graphene~\cite{Novoselov and Grigorieva,Manzeli_Kis}, have begun a new field of research because of their distinctive physical properties, hence, their different and interesting applications in industrial sectors like optoelectronics, magnetic recording, energy, thermo-electrics, etc.~\cite{Li_Zhu,Deng_Bao,Mahapatra_2017,Zhang_Ciesielski} that in turn gives rise to the new technologies. This magnetism is coming due to the spins. For the fundamental understanding the Ising model, a historical model, is studied in physics that provides a simplified framework to investigate complex spin interactions in 2D and 3D materials offering valuable insights of the phenomena. Here we are focusing on 2D systems because of their special properties. The range of interactions has a role of great importance, for example, new long-range interactions have been recently studied in~\cite{li_2024} and the references mentioned therein. The variation of the strength of interaction alters the detailed balance condition of the system, affecting the system's overall thermodynamic behavior. This is one of our interests in this study.\\
Let's have some account of an engrossing phenomenon--the detailed balance condition, crucial in Metropolis-Hastings' algorithm, which states that at equilibrium, all elementary processes should be equilibrated by their reverse processes. In simple terms, the probability of the system transitioning from state $A$ to state $B$ is the same as the probability of transitioning from state B to state A. In the context of the Ising model, which describes the magnetic spins on a lattice, the detailed balance condition is of great importance for sampling the configuration space according to the Boltzmann distribution. A random move is accepted or rejected based on the change in energy, following the Metropolis algorithm. The Metropolis criteria are essential to ensure that the detailed balance condition is satisfied~\cite{landau_binder_2014}.\\
The detailed balance condition states that for any two states $\sigma$ and $\sigma^{\prime}$ of the system, the transition rates $\mathrm{W}\left(\sigma \rightarrow \sigma^{\prime}\right)$ and $W\left(\sigma^{\prime} \rightarrow \sigma\right)$ must satisfy:
\begin{equation}
	\frac{\mathrm{P}\left(\sigma^{\prime}\right) \mathrm{W}\left(\sigma^{\prime} \rightarrow \sigma\right)}{\mathrm{P}(\sigma) \mathrm{W}\left(\sigma \rightarrow \sigma^{\prime}\right)}=1 
	\label{eq_2}
\end{equation}
where $P(\sigma)$ denotes the equilibrium probability distribution of state $\sigma$. For a system in thermal equilibrium at temperature $T$, the probability $(\sigma)$ is given by the Boltzmann distribution:
\begin{equation}
	P(\sigma)=\frac{e^{-\beta H(\sigma)}}{\mathrm{Z}} 
	\label{eq_3}
\end{equation}
where $\beta=\frac{1}{k_{B} T}$ ($k_{B}$ is Boltzmann constant), $H(\sigma)$ is the energy of state $\sigma$, and $Z$ is the partition function:
\begin{equation}
	\mathrm{Z}=\sum_{\sigma} \mathrm{e}^{-\beta \mathrm{H}(\sigma)} 
	\label{eq_4}
\end{equation}
To ensure detailed balance, the transition rates should be chosen such that:
\begin{equation}
	\frac{\mathrm{e}^{-\beta \mathrm{H}\left(\sigma^{\prime}\right)} \mathrm{W}\left(\sigma^{\prime} \rightarrow \sigma\right)}{\mathrm{e}^{-\beta \mathrm{H}(\sigma)} \mathrm{W}\left(\sigma \rightarrow \sigma^{\prime}\right)}=1 
	\label{eq_5}
\end{equation}
which simplifies to:
\begin{equation}
	\frac{\mathrm{W}\left(\sigma^{\prime} \rightarrow \sigma\right)}{\mathrm{W}\left(\sigma \rightarrow \sigma^{\prime}\right)}=\mathrm{e}^{\beta\left[\mathrm{H}\left(\sigma^{\prime}\right)-\mathrm{H}(\sigma)\right]} 
	\label{eq_6}
\end{equation}
One common choice for the transition rates that satisfies the detailed balance condition is the Metropolis-Hastings algorithm, where the transition rate from state $\sigma$ to $\sigma^{\prime}$ is given by:
\begin{equation}
	\mathrm{W}\left(\sigma \rightarrow \sigma^{\prime}\right)=\min \left(1, e^{\beta\left[\mathrm{H}\left(\sigma^{\prime}\right)-\mathrm{H}(\sigma)\right]}\right) 
	\label{eq_7}
\end{equation}
Critical slowing down~\cite{landau_binder_2014} shows auto-correlation time blows up. Since the Metropolis algorithm flips a single random spin at a time to dismantle clusters spread throughout the lattice, this process is a huge time-consuming process in simulations. The correlation length is the characteristic length scale over which the spins of the system are correlated. Near the transition temperature, the correlation length diverges, indicating that correlations extend over long distances. Additionally, the correlation time, which is the characteristic time scale over which spins in the system are correlated, also increases. At the critical point, where there is a phase transition between the ordered (ferromagnetic) phase and the disordered (paramagnetic) phase, correlations become stronger and extend over long distances. Due to this, the single-flip method offered by the Metropolis algorithm leads to a highly time-consuming process known as critical slowing down~\cite{wolff_1989, landau_binder_2014}.\\
The cluster flipping method offered by the Wolff algorithm ~\cite{wolff_1989} assists in de-correlating the system in significantly less time. This can be realized from the Eq.\eqref{eq_8}  that gives the relation between the correlation time $(\tau)$ and the correlation length ($\xi$).
\begin{equation}
	\tau \approx \xi^{z} 
	\label{eq_8}
\end{equation}
Here, $z$ represents the dynamic exponent. The value of this dynamic exponent is approximately $\approx 2.17$ for the Metropolis algorithm and nearly zero for the Wolff algorithm~\cite{wolff_1989,wansleben_landau_1991}.\\
The systems we are studying have limited size, unlike the realistic models that ideally extend to infinite size $(L \rightarrow \infty)$. Therefore, boundary conditions need to be taken into consideration, and for that, we are employing the usual approach of using periodic boundary conditions. This implies that the topmost and rightmost layers are considered neighbors, as are the bottom-most and leftmost layers.\\
\begin{figure}[ht] 
	\centering
	\includegraphics[width=16cm,height=7cm]{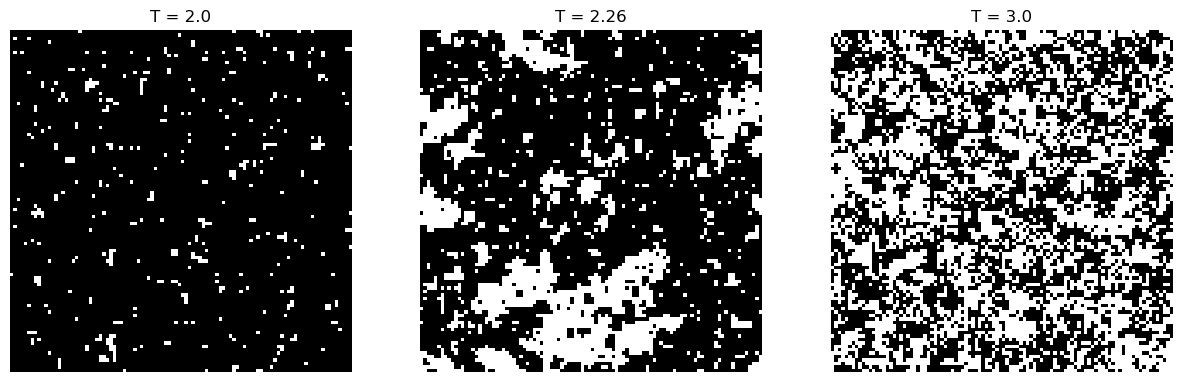}
	\includegraphics[width=16cm,height=7cm]{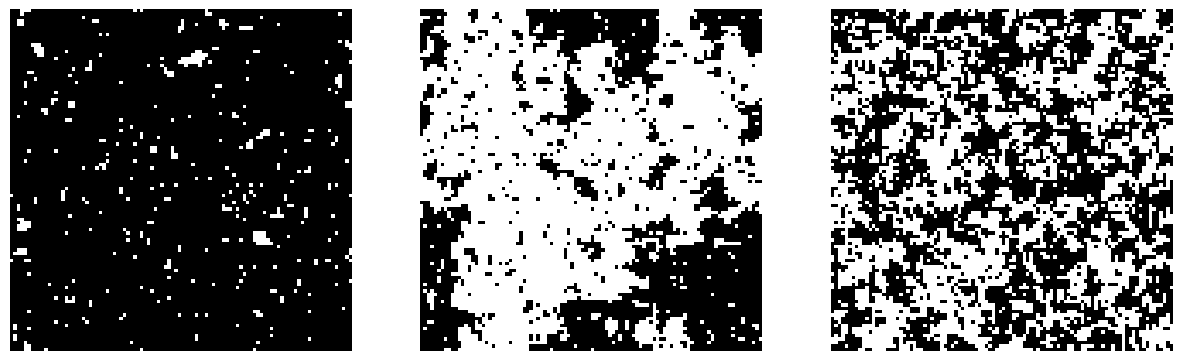}
	\caption{Snapshot of the Monte-Carlo simulation of isotropic Ising model using Wolff algorithm (first row) and the MA (second row) at three (2.0, 2.26, and 3.0) temperatures (Black and white dots here represent the +1 and -1 states of the model, video file:~\cite{youtube}).}
	\label{Figure_2}
\end{figure}
Fig.\ref{Figure_2} presents snapshots of the simulation at three different temperatures of the isotropic Ising model: 2.0 (below the transition temperature), 2.26 (near the transition temperature), and 3.0 (above the transition temperature). Black and white dots here represent the +1 and -1 states of the model. The first row displays the results for the Wolff algorithm, while the second row shows the lattice of the Metropolis algorithm. These snapshots were captured after $10^{3}$ Monte Carlo steps for $ 100 \times 100$ lattice. For temperatures near the critical temperature, the Wolff algorithm reaches equilibrium near $10^{3}$ steps, whereas the Metropolis algorithm lags. A video demonstrating these flips has been created and an online video can be seen by clicking the link in~\cite{youtube} From the video, it can be observed that even at a temperature of 2.0 (below the transition temperature) and especially at T = 2.26 (near to the transition temperature), the Metropolis algorithm takes larger number of Monte-Carlo steps for equilibration, while the Wolff algorithm shows the transition of the system in lesser number of Monte-Carlo steps. At T=3.0 (above transition temperature) the system is not transitioning from the ferromagnetic to the paramagnetic state. The phase transition is not possible above the transition temperature because the thermal fluctuations dominate the system.\\
Our main results include magnetization, energy, specific heat, magnetic entropy, and Binder cumulant versus temperature at various magnetic field values. Additionally, we present the study of the Magnetocaloric effect for both the isotropic Ising model anisotropic Ising model and different anisotropic Ising model cases. Transition temperatures for the Isotropic Ising model and four different variants of the anisotropic Ising model were determined through Monte Carlo simulations using both the Metropolis and Wolff algorithms. The four different variants of the anisotropic Ising model involve five distinct values of $\frac{J_{y}}{J_x}$, namely $0.5, 0.75, 1.0, 1.25$, and 1.5. Transition temperatures of these models were identified using Binder cumulant graph plotting for five different values of $N$, the lattice size. The values of $N$ considered in this study are 10, 20, 30, 40, and 50.\\
The magneto-caloric effect (MCE) is an interesting topic for the discovery and development of new materials with enhanced magnetocaloric properties. Contrary to traditional methods of gas compression and expansion cycles for refrigeration, for example, which often use environmentally harmful refrigerants, magnetocaloric refrigeration has the potential to be a more efficient source of energy and friendly to the environment. Magnetocaloric refrigeration can provide targeted cooling with high precision which has high industrial uses. Magnetocaloric technologies can be integrated with renewable energy sources. The MCE comes with some scientific and technological challenges that drive research and innovation. Though MCE promises many things, understanding the fundamental physics behind the effect, optimizing material properties, and designing efficient magnetocaloric devices are table-top topics in research. These challenges stimulate advancements in materials science, thermodynamics, and engineering. The investigation of magnetic properties and magnetocaloric effect has been an interesting topic for a long time. Even nowadays researchers are exploring these properties in different systems using Monte Carlo and other methods. The magnetocaloric effect using Monte Carlo simulations has been discussed for 1D, 2D, and 3D ferrimagnetic systems has been discussed in~\cite{stanica_2015}. Magnetic properties and magnetocaloric effect of an Ising-type polyhedral chain have been discussed in ~\cite{yang_2022}, on 2D core-shell Ising system in~\cite{ghazrani_2024}, Ising model with RKKY interaction in~\cite{Lubomira_2022}, 3D Ising model with new long-range interaction in~\cite{li_2024} Further discussion about MCE is in the magneto-caloric effect section later.\\
After the introduction in the above section \ref{S_1}, the rest of the paper is framed as follows. First, we are presenting an overview of the Metropolis and Wolff algorithm in section \ref{S_2}. Afterward, a discussion on symmetry, order parameters, and phase transitions is given in section \ref{S_3}. In the next section, we are moving to results and analysis of the below-mentioned thermodynamic quantities (Eq.\eqref{eq_11}) to (\ref{eq_14})) for anisotropic Ising model and anisotropic Ising model using Metropolis algorithm and Wolff algorithm, followed by the study of transition temperatures using Binder cumulant using Wolff algorithm, in section \ref{S_4}. Subsequently, we are presenting the study of the thermodynamic and magnetic quantities ((Eq.\eqref{eq_11}) to \eqref{eq_15}) at varying magnetic fields in section \ref{S_5}. Thereafter, we are presenting the hysteresis loop survey for the Ising model for different Isotropic cases in section \ref{S_6}. Finally, we go to the magnetocaloric effect in section \ref{S_7}. And the conclusion of the paper in the last section \ref{S_8}.
\section{The Metropolis and Wolff algorithm}
\label{S_2}
The Metropolis and the Wolff algorithm have been well studied in the literature for example in~\cite{kalos_whitlock_1986, metropolis_1953, luijten_2006, tierney_1994, Warburg_1881}. Here we are giving the overview of the step-by-step process of how these algorithms work~\cite{wolff_1989, stanica_2015}.\\
\subsection{Metropolis-Hastings Algorithm for the Ising Model}
\label{MHA}
\begin{enumerate}
	\setcounter{enumi}{0}
	\item Initialize the system
	\begin{itemize}
		\item Lattice: Consider a 2D lattice of size $N \times N$ where each site $\langle i, j\rangle$ can have a spin which can be either +1 or -1.
		\item Initial Configuration: Initialize the spins randomly, i.e., each $\langle i, j\rangle$ can be +1 or -1 with equal probability.
		\item Define the Hamiltonian. The Hamiltonian for the Ising model without an external magnetic field is given by Eq.\eqref{eq_1}.
		\item The sum runs over all nearest-neighbor spins.
	\end{itemize}
	\item Set parameters:
	\begin{itemize}
		\item Temperature $(T)$
		\item Boltzmann constant $(k_{B})$
		\item Compute $\beta=\frac{1}{k_{B} T}$
	\end{itemize}
	\item Monte Carlo steps: Perform the following steps repeatedly to evolve the system:
	\begin{itemize}
		\item For each Monte Carlo step repeat the fallowing two steps:
		\begin{itemize}
			\item Select a spin: Randomly select a spin $\langle i, j\rangle$ on the lattice.
			\item Calculate energy change: Compute the change in energy $\Delta E$ if the selected spin is flipped:
			\begin{equation}
				\Delta E=2 J s_{i, j} \sum_{k, l} S_{k, l} 
				\label{eq_9}
			\end{equation}
			where the sum runs over the nearest neighbors $(\mathrm{k}, \mathrm{l})$.
		\end{itemize}
	\end{itemize}
 	\item Metropolis criterion: Decide whether to accept the spin flip:
 	 	\begin{itemize}
 		\item If $\Delta E \leq 0$, accept the flip (i.e., flip the spin).
 		\item If $\Delta E>0$, accept the flip with probability $e^{-\beta \Delta E}$.
 		\end{itemize}
	\item Update the system
	\begin{itemize}
		\item If the flip is accepted, update the spin configuration.
		\item If the flip is not accepted, keep the spin configuration unchanged.
	\end{itemize}
	\item Repeat the Monte Carlo steps for a large number of iterations to ensure the system reaches equilibrium.
\end{enumerate}.
\subsection{Wolff Algorithm for the Ising Model}
\label{WA}
Step 1 to 3 are same as for Metropolis-Hastings' algorithm.
\begin{enumerate}
	\setcounter{enumi}{3}
	\item Perform the following steps, for each Monte Carlo step, repeatedly to evolve the system:
	\begin{itemize}
		\item Random seed selection: Randomly select a seed spin $\mathrm{s}_{\mathrm{i}, \mathrm{j}}$ on the lattice.
	\end{itemize}
	\item Cluster growth:
	\begin{itemize}
		\item Initialize the cluster with the seed spin.
		\item Flip the seed spin.
		\item Use a stack or queue to keep track of spins to be checked for addition to the cluster.
	\end{itemize}
	\item Cluster formation:
	\begin{itemize}
		\item For each spin in the cluster, consider its neighboring spins.
		\item For each neighboring spin $\mathrm{s}_{\mathrm{k}, \mathrm{l}}$ that is aligned (same sign) with the current spin:
		\item Calculate the probability 
		\begin{equation}
			\quad P_{add}=1-e^{2 \beta J} 
			\label{eq_10}
		\end{equation}
		\item With probability $P_{a d d}$ the neighboring spin to the cluster and flip it.
		\item Continue this process until no more spins can be added to the cluster.
	\end{itemize}
\end{enumerate}
\section{Symmetry, Order Parameter and Phase transitions}
\label{S_3}
One of the fundamental concepts in condensed matter physics, in addition to other branches of physics, is spontaneous symmetry breaking~\cite{chaikin_lubensky}. This phenomenon occurs when a system that is initially symmetric transforms, losing its symmetry without any explicit external cause. Symmetry refers to invariance under a set of transformations. Spontaneous symmetry breaking happens when the underlying laws governing a system are symmetric, but the system itself settles into a state that lacks this symmetry. The symmetry of the system is "broken" by the dynamics of the system itself, not by external influences. In the case of the Ising model, at high temperatures, spins in a ferromagnetic material are disordered, and the system is symmetric under spin flips (up to down and vice versa). Below the transition temperature, spins align in one direction, breaking the spin-flip symmetry. The system chooses a direction of magnetization, either up or down, despite the underlying symmetric laws.\\
An order parameter is a measure that describes the degree of symmetry breaking. In ferromagnetism, magnetization ($M$) serves as an order parameter. In the symmetric phase (high temperature), $M=0$. In the broken symmetry phase (low temperature), $M \neq 0$. Symmetry breaking leads to multiple equivalent ground states. Choosing one of these states breaks the symmetry. Near the phase transition where symmetry breaking occurs, systems exhibit critical phenomena such as diverging correlation lengths and susceptibilities. Spontaneous symmetry breaking is a profound concept that explains the emergence of order and structure in various physical systems despite symmetric underlying laws. It plays a crucial role in understanding phase transitions, the behavior of fundamental particles, and the evolution of the universe.\\
Magnetization per unit spin ($M$) in the Ising model can take two values: +1 or -1, representing whether all spins are aligned in the upward or downward direction. In the absence of an external magnetic field, the system exhibits spontaneous magnetization, which means that $M$ taking a value of +1 or -1 is equally probable. However, over time, due to thermal fluctuations, $M$ can transition from +1 to -1 and vice versa.  All of this is possible below the transition temperature of the system. Below the critical temperature, the system can get trapped in a so-called meta-stable state, where the system remains in a particular state for a sufficiently long time before switching to the other one. In the Ising model, Monte Carlo (MC) is used for random updating of the spins. This way, the dynamics of the Ising model through MC allow the transition of the system between these magnetization states. Fig.\ref{Figure_3} shows the transition of the model between two magnetization states that is between +1 and -1 states.
\section{Results and Analysis}
\label{S_4}
\begin{figure}[ht] 
	\centering
	\includegraphics[width=16cm,height=4.5cm, trim= 0 307 0 20, clip]{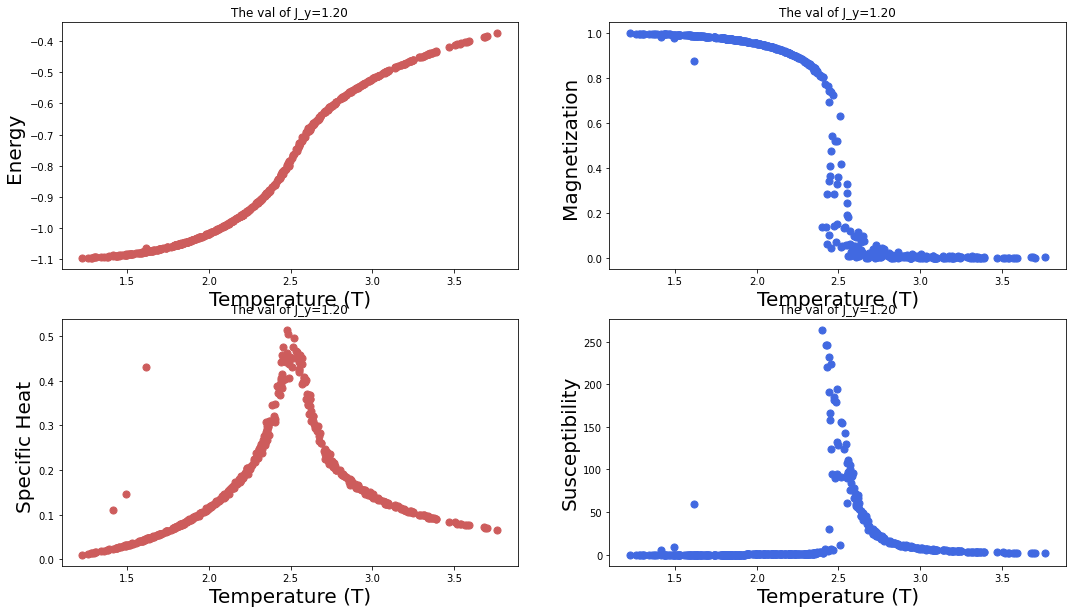}   
	\includegraphics[width=16cm,height=4.5cm, trim= 0 0 0 319, clip]{fig_3_a_MECX_J_y_1.2_32x32.png}\\
	(a)\\
	\includegraphics[width=16cm,height=9cm]{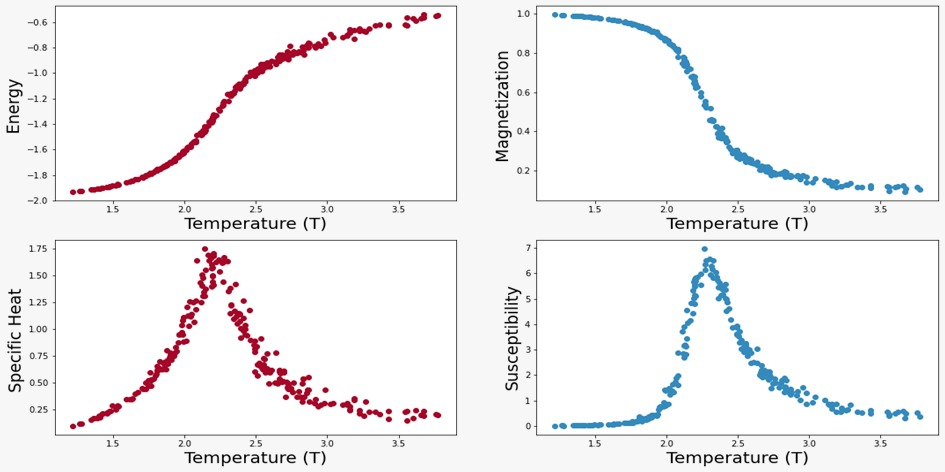}\\
	(b)\\
	\caption{Energy Magnetization, Specific heat and Susceptibility graph for (a) temperature higher than transition temperature for $\frac{J_y}{J_x} = 1.2$ and $32 \times 32$ lattice (b) temperature near to transition temperature for $\frac{J_y}{J_x} = 1.0$  and $16 \times 16$ lattice.}
	\label{Figure_4}
\end{figure}

\begin{figure}[ht] 
	\centering
	\includegraphics[width=8cm,height=8cm]{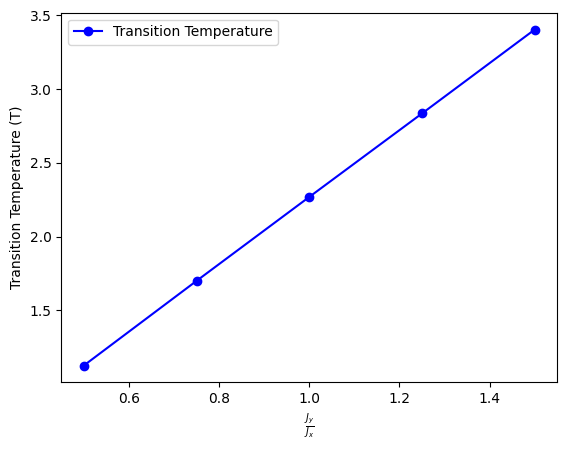}
	\caption{Plot for showing the variation of transition temperatures for different $\frac{J_y}{J_x}$ values.}
	\label{Figure_6}
\end{figure}

The different physical quantities we are studying here are: magnetization (M), energy (E), specific heat (C), susceptibility ($\chi$), magnetic entropy($S_M$), and Binder cumulant ($U_L$). The formulas for these quantities are:
\begin{gather}
	\langle\mathrm{M}\rangle=\frac{1}{\mathrm{N}} \sum_{\mathrm{i}}^{\mathrm{N}} \mathrm{s}_{\mathrm{i}} \label{eq_11}\\	
	\langle\mathrm{E}\rangle=\frac{1}{\mathrm{N}} \sum_{\mathrm{i}}^{\mathrm{N}} \mathrm{H}_{\mathrm{i}} \label{eq_12}\\
	\mathrm{C}=\frac{\beta}{\mathrm{T}}\left[\left\langle\mathrm{E}^{2}\right\rangle-\langle\mathrm{E}\rangle^{2}\right]\label{eq_13}\\
	\chi=\beta\left[\left\langle\mathrm{M}^{2}\right\rangle-\langle\mathrm{M}\rangle^{2}\right] \label{eq_14}\\
	\mathrm{S_M}(\mathrm{T},\mathrm{h})=\int_{0}^{\mathrm{T}} \frac{\mathrm{C}}{\mathrm{T}} \mathrm{dT} \label{eq_15}\\
	\mathrm{U_L}=1-\frac{\left\langle\mathrm{M}^{4}\right\rangle}{3\left\langle\mathrm{M}^{2}\right\rangle^{2}} \label{eq_16}
\end{gather}
$N$ is the lattice size, $\beta$ is inverse temperature and $T$ is the temperature.\\
To ensure that the quantities measured represent the steady-state properties of the system, the measurements are taken when the system is in equilibrium. The steady state, or equilibrium state, is the state where the properties of the system no longer change with time. The system begins with a random initial configuration and is allowed to evolve by the MC method through the Metropolis algorithm to reach equilibrium. After equilibration, the system enters the measurement phase to evaluate the statistical properties (such as Eq.\eqref{eq_11} to Eq.\eqref{eq_16}).\\
\begin{figure}[ht] 
	\centering
	\includegraphics[width=16cm,height=7cm]{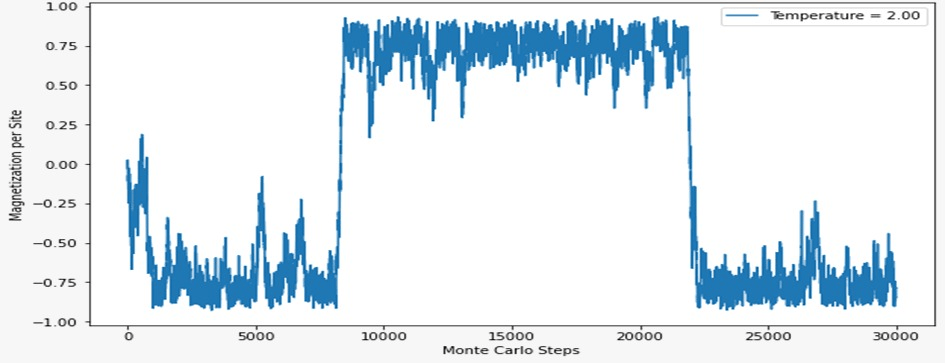}
	(a)
	\includegraphics[width=16cm,height=7cm, trim= 0 0 0 17.7, clip]{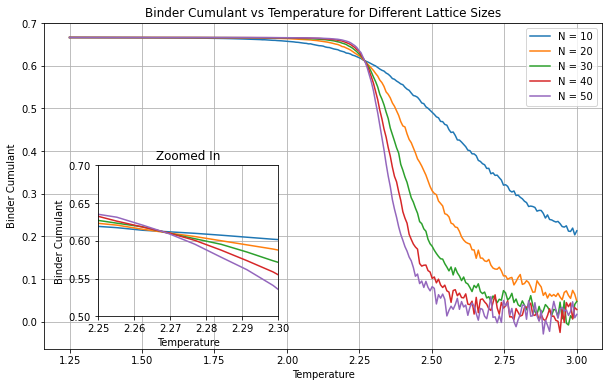}
	(b).\\
	\caption{ Magnetization vs Monte-Carlo steps (a) two magnetization states and (b) Binder cumulant graph showing transition temperature for isotropic Ising model.}
	\label{Figure_3} 
\end{figure}
The Binder cumulant $(U_L)$, a dimensionless quantity, is used to determine the critical temperatures in statistical physics. $U_L$ is given by Eq.\eqref{eq_16}. To benchmark our codes, we are finding the transition temperature of the model for the isotropic case. Near the transition temperature, $U_L$ demonstrates a peculiar behavior, which is used to pinpoint the transition temperature. The point where these Binder cumulant curve lines intersect each other is considered the transition temperature. The point where these $U_L$ curves intersect can be tracked with more accuracy (in terms of higher decimals) as we increase the number of MC steps. Another factor that can be used to obtain a more precise value of the transition temperature is increasing the size of the lattice. Additionally, in the plots of the thermodynamic quantities, the steepness of the curves of the statistical quantities increases as we move to higher lattice sizes, indicating the exact transition temperature more profoundly. (This can be observed in Fig.\ref{Figure_4}(a), which has been plotted for a lattice size of 32, and Fig.\ref{Figure_4}(b), which has been plotted for a lattice size of 16.) The isotropic case means $\frac{J_y}{J_x}=1.0$, and the plot is shown in Fig.\ref{Figure_3}. For this case, we have found the transition temperature($T_c$) to be approximately 2.267, which is quite close to Onsager's exact value calculation~\cite{onsager_1944} for the 2D Ising model. The exact value is given by:\\
\begin{equation}
	T_{c}=\frac{2 J}{\ln (1+\sqrt{2})} \approx 2.269 \mathrm{J} 
	\label{eq_17}
\end{equation}
This curve is plotted for five different values of lattice size, namely 10, 20, 30, 40, and 50. These plots are drawn for $2^{11}$ equilibration steps and $2^{15}$ Monte-Carlo steps. The fluctuations in the $U$ plot at higher temperatures are due to an increase in thermal fluctuations. The Wolff algorithm is used for the calculation of magnetization and hence the Binder cumulant.\\
The changes observed in the graphs of the calculation of the thermodynamic physical quantities like energy, magnetization, specific heat, and susceptibility can be observed in the plots shown in Fig.\ref{Figure_4}(a). In these plots, the value of $\frac{J_y}{J_x}$ is 1.20. The peaks observed near the transition temperature in the specific heat and susceptibility graphs have shifted towards higher temperatures, nearly 2.5. Similarly, an increase or decrease in the energy or magnetization graph is also observed at the same temperature. In Fig.\ref{Figure_4}(b), the same thermodynamic quantities are plotted but for $\frac{J_y}{J_x}= 1.0$. The shift, which is towards higher temperature for higher $\frac{J_y}{J_x}$ values, is towards lower temperature for lower $\frac{J_y}{J_x}$ values. The number of MC steps taken for the plots in Fig.\ref{Figure_4} (a) and (b) is the same.\\
\begin{table}
\begin{center}
	\caption{Different $\frac{J_y}{J_x}$ values and their corresponding transition temperatures.}
	\begin{tabular}{|c|c|c|}
		\hline
		Figure & $\frac{J_y}{J_x}$ value & \begin{tabular}{c}
			Transition \\
			temperature \\
		\end{tabular} \\
		\hline
		$5(\mathrm{a})$ & 0.50 & $1.126 \mp 0.001$ \\
		\hline
		$5(\mathrm{b})$ & 0.75 & $1.700 \mp 0.001$ \\
		\hline
		$3(\mathrm{b})$ & 1.00 & $2.269 \mp 0.001$ \\
		\hline
		$5(\mathrm{c})$ & 1.25 & $2.835 \mp 0.001$ \\
		\hline
		$5(\mathrm{d})$ & 1.50 & $3.404 \mp 0.001$ \\
		\hline
	\end{tabular}
	\label{Table_1}
\end{center}
\end{table}
Fig.\ref{Figure_5} provides the Binder cumulant graphs for four different values of $\frac{J_y}{J_x} \ne 1$. The point where these Binder cumulant curves intersect each other gives the transition temperature of the model. Therefore, the transition temperatures for anisotropic Ising models for four different values of $\frac{J_y}{J_x}$ are found. The different $\frac{J_y}{J_x}$ values and their transition temperatures are shown in Table \ref{Table_1}. Finally, in Fig.\ref{Figure_6}, we are showing the variation of transition temperatures versus different $\frac{J_y}{J_x}$ values. The Binder cumulants are calculated using the Wolff algorithm.\\
\begin{figure}[ht] 
	\centering
	\includegraphics[width=8cm,height=4.5cm, trim= 0 0 0 17.85, clip]{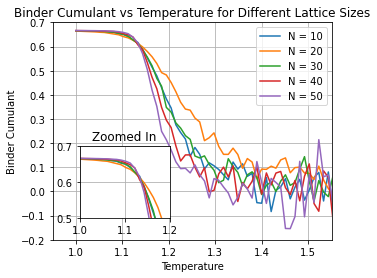}(a)
	\includegraphics[width=8cm,height=4.5cm, trim= 0 0 0 17.85, clip]{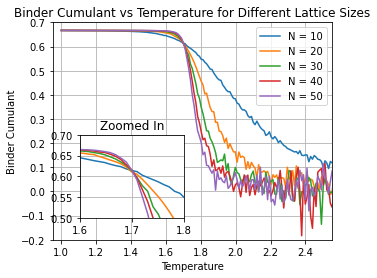}(b)
	\includegraphics[width=8cm,height=4.5cm, trim= 0 0 0 17.85, clip]{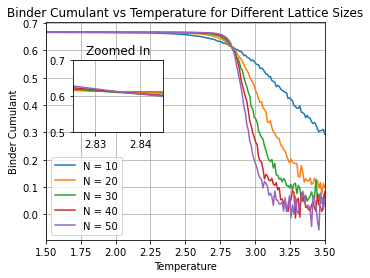}(c)
	\includegraphics[width=8cm,height=4.5cm, trim= 0 0 0 17.85, clip]{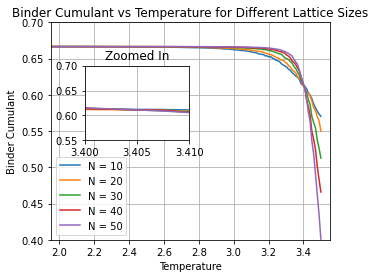}(d).\\
	\caption{Binder cumulant graph for different $J_y$ values showing their transition temperatures. (a) $\frac{J_y}{J_x}=0.50$ (b) $\frac{J_y}{J_x}=0.75$ (c) $\frac{J_y}{J_x}=1.25$ (d) $\frac{J_y}{J_x}=1.5$. Their corresponding transition temperatures are given in table \ref{Table_1}(The graph for $\frac{J_y}{J_x}=1.0$) is given in Fig.\ref{Figure_3}(b).}
	\label{Figure_5}
\end{figure}
\section {Effects of External Magnetic Field}
\label{S_5}
\begin{figure}[ht] 
	\centering
	\includegraphics[width=16cm,height=5.5cm, trim= 0 0 0 360, clip]{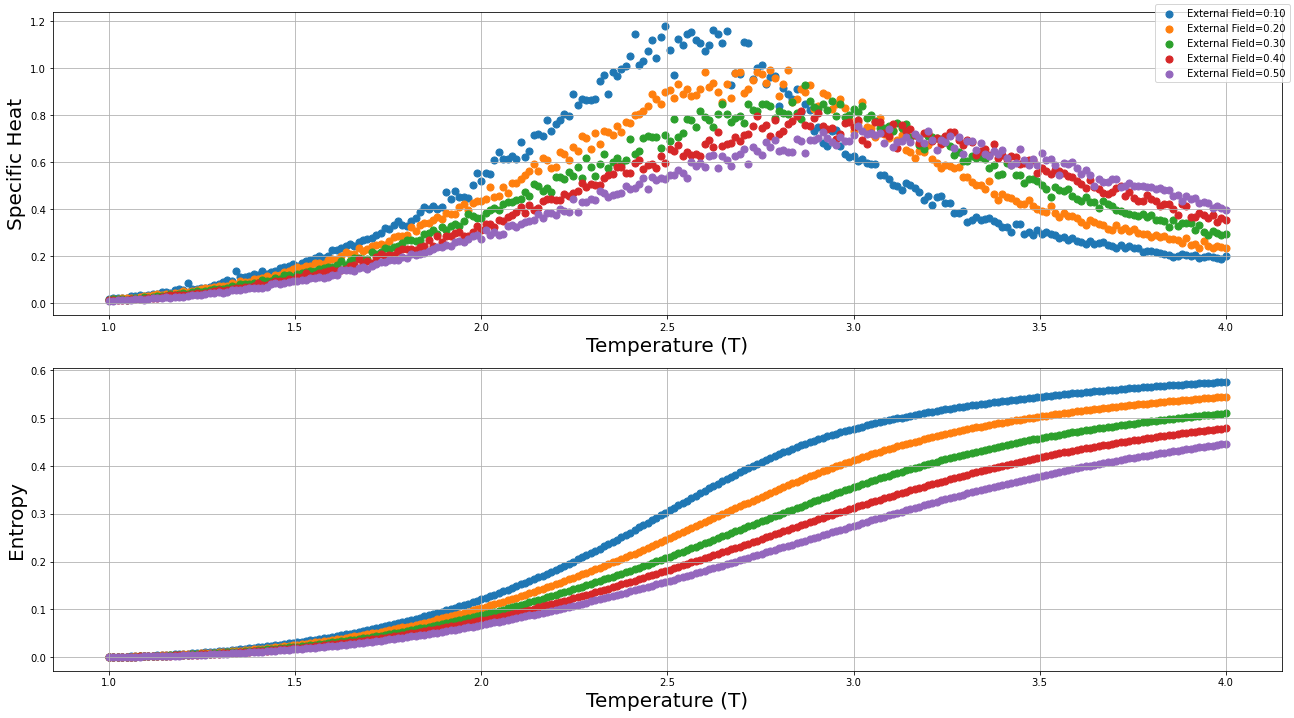}
	\includegraphics[width=16cm,height=10cm]{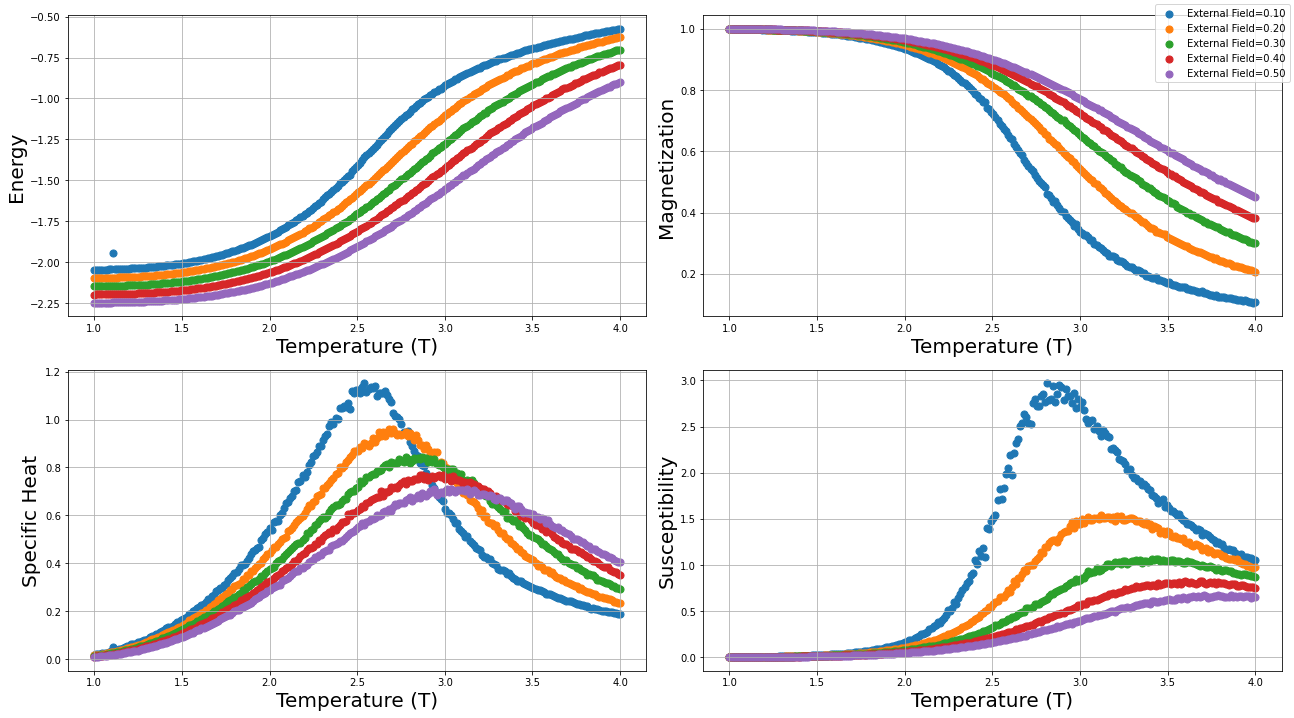}.\\
	\caption{Variation of thermodynamic quantities with respect to temperature at different values of externally applied magnetic field(h).}
	\label{Figure_7}
\end{figure}

The Ising Hamiltonian includes a second term representing an applied magnetic field effect. Studying different thermodynamic quantities at various externally applied magnetic field values is also our interest. In Eq.\eqref{eq_1}, the second term $h_{i}$ represents the applied uniform magnetic field, where $h_{j}=h$. We are conducting a study of different statistical quantities at five different values of $h: 0.1,0.2,0.3,0.4$, and $0.5$ in units of $J$. Plots for energy, magnetization, specific heat, susceptibility, and magnetic entropy with respect to temperature for the aforementioned different values of externally applied magnetic fields are generated using the Metropolis Algorithm by running for $2^{16}$ Monte Carlo steps, and are shown in Fig.\ref{Figure_7}.\\
In the presence of an external magnetic field, the spins of the system align with the field. Magnetization gradually decreases with the increase in temperature, but due to the applied field, it remains non-zero. The energy of the system is lower due to the alignment of the spins along the field, causing a decrease in energy. With the temperature rise, the energy increases due to thermal fluctuations, but a slackening in the rate of increase is observed. The specific heat and susceptibility peaks broaden in comparison to the non-zero field case, in the presence of the magnetic fields. From the results, it can be seen that the temperature at which the energy, magnetization, and magnetic entropy plots start saturating increases with the increase in the values of the externally applied magnetic field. Additionally, the temperatures at which the specific heat and susceptibility graphs show peaks, indicating the transition temperatures, also move towards higher temperatures. Therefore, the transition temperatures are increasing.
\section{The hysteresis loops}
\label{S_6}
\begin{figure}[ht] 
	\centering
	\includegraphics[width=14cm,height=8.5cm, trim= 0 0 0 19, clip]{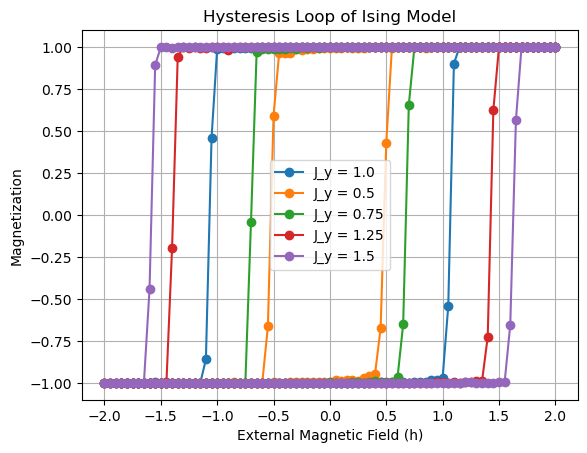}.\\
	\caption{Hysteresis loop of isotropic and anisotropic Ising model for different $\frac{J_y}{J_x}$(shown as J$\_$y) values.}
	\label{Figure_8}
\end{figure}

The exhibition of the hysteresis loop is observed when the magnetic field is varied cyclically, demonstrating the delay in the alignment of spins with the field, and hence the lag between the applied magnetic field and the response of magnetization of the system. The hysteresis loop of the isotropic Ising model is also studied. Fig.\ref{Figure_8} shows the hysteresis loop of the Ising model for 5 different $\frac{J_y}{J_x}$ values: $1.0,1.5,1.25,0.5$, and $0.75$. The range of the change in the magnetic field values(h) is taken from -2 to 2, in units of J. The Monte Carlo steps taken in this case are $5 \times 10^{3}$ steps. The provided graph shows hysteresis cycles for different values of $\frac{J_y}{J_x}$ in an Ising model. The width of the loop increases with the increase in the $\frac{J_y}{J_x}$ value.\\
\section{Magnetocaloric Effect}
\label{S_7}
Another important thermodynamic quantity of significant industrial usage is the magnetocaloric effect (MCE). Since the concept of MCE was first presented, it has been of interest to the both theoretical and experimental physics community, from at least two points of view--magnetic refrigeration and phase transitions. Since then, many studies have been made to discuss various aspects of MCE~\cite{stanica_2015,yang_2022,li_2024,Lubomira_2022,li_dpt_2024,ghazrani_2024,li_dpt_2024,Tishin_2003} and the references therein. The Ising model, a basic model highly studied in the physics community for its magnetic properties, can be used to further understand MCE. The isothermal magnetic entropy changes $\Delta S_{M}$ is a key parameter characterizing MCE, calculated by the following Eq.:
\begin{equation}
	\Delta \mathrm{S}_{\mathrm{M}}=\int_{0}^{\mathrm{h}}\left(\frac{\partial \mathrm{M}}{\partial\mathrm{T}}\right)_{\mathrm{h}^{\prime}} \mathrm{dh}^{\prime} 
	\label{eq_18}
\end{equation}
Here $h$ is the magnetic field.\\
\begin{figure}[ht] 
	\centering
	\includegraphics[width=8cm,height=4.5cm]{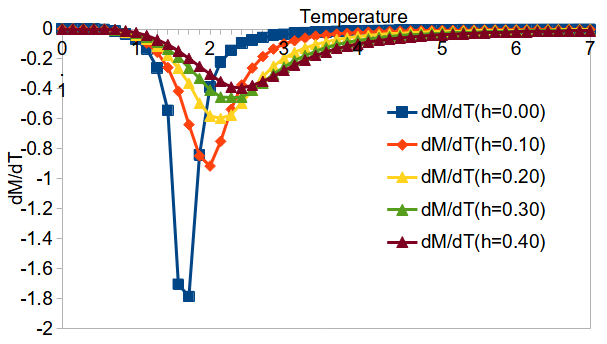}
	\includegraphics[width=8cm,height=4.5cm]{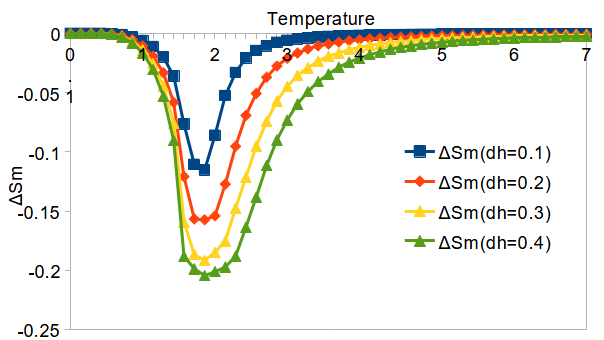}(a)
	\includegraphics[width=8cm,height=4.5cm]{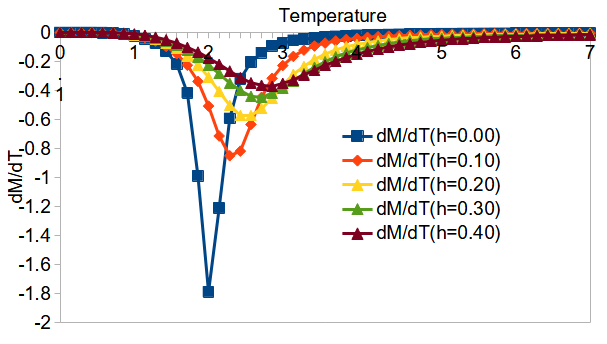}
	\includegraphics[width=8cm,height=4.5cm]{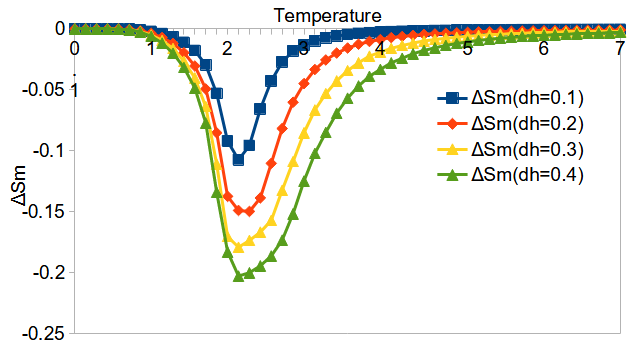}(b)
	\includegraphics[width=8cm,height=4.5cm]{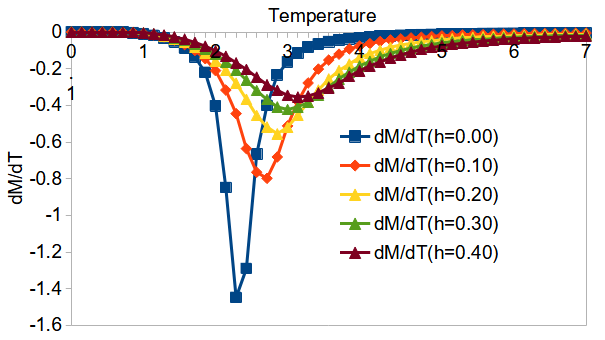}
	\includegraphics[width=8cm,height=4.5cm]{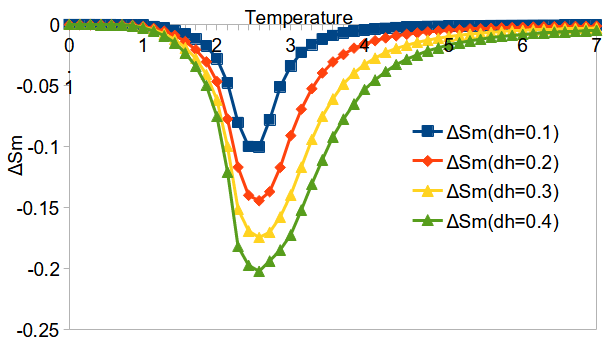}(c)
	\includegraphics[width=8cm,height=4.5cm]{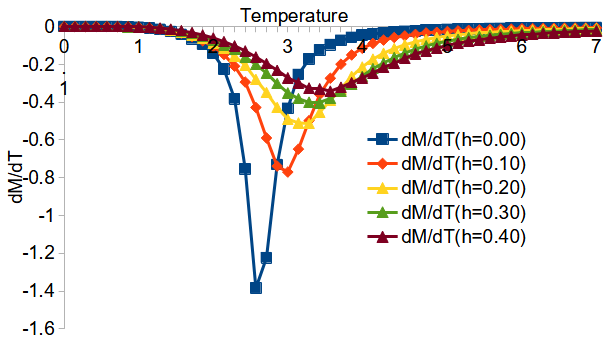}
	\includegraphics[width=8cm,height=4.5cm]{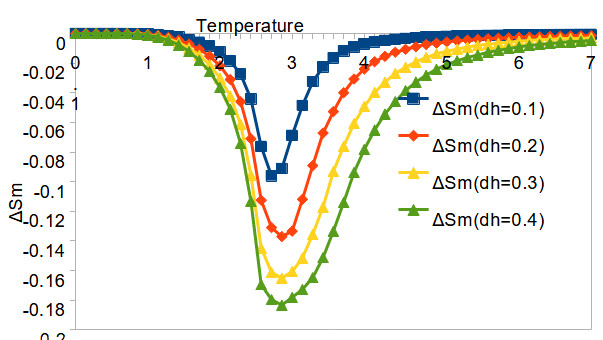}(d)
	\includegraphics[width=8cm,height=4.5cm]{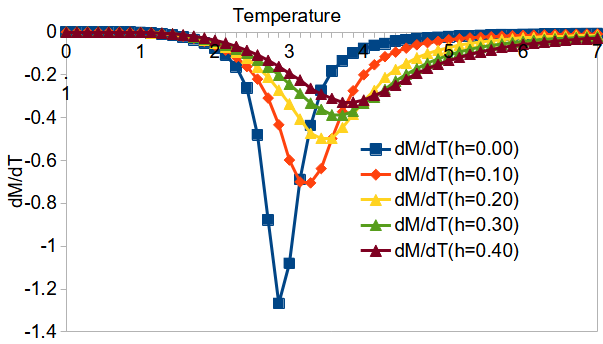}
	\includegraphics[width=8cm,height=4.5cm]{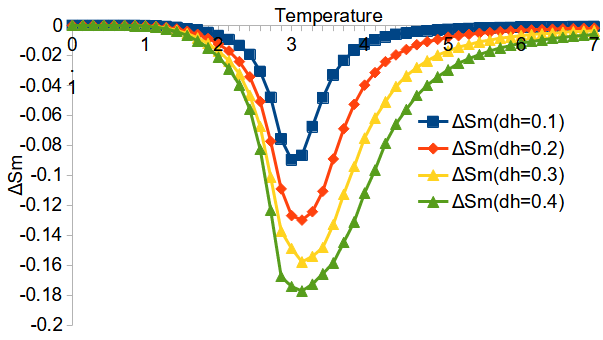}(e)\\
	\caption{The left column is $\frac{\partial M}{\partial T}$(shown as dM/dT) and the right column is change in magnetic entropy $\left(\Delta \mathrm{S}_{\mathrm{M}}\right)$(shown as $\Delta Sm$) of different cases for the $\frac{J_y}{J_x}$ values of (a)0.5, (b)0.75, (c) 1.0 (d)1.25, (e)1.5 with respect to temperature.}
	\label{Figure_9}
\end{figure}
The phenomenon of the MCE works as follows: when an external magnetic field is applied adiabatically, all the magnetic moments (spins) become oriented in a single direction, leading to a decrease in the total magnetic entropy of the material. As a result, the material heats up. Upon removal of the magnetic field, the material returns to its original state, absorbing all the heat. This is necessary for the total magnetic entropy to remain constant, requiring the magnetic entropy of the crystal lattice to increase. When the magnetic field is removed adiabatically, the material emits heat, causing it to cool down. The MCE is more prominent near the transition temperature of the material.\\
The derivative of magnetization to the temperature ($\frac{\partial M}{\partial T}$) provides insights into how the magnetization of the system responds to changes in temperature. Near the phase transition point, the system exhibits a significant change, manifesting as a peak similar to specific heat and susceptibility but in the negative direction. This negative peak diminishes with an increase in the external magnetic field. The MCE is computed as the isothermal magnetic entropy change ( $\Delta S$ or $\Delta S_{M}$ ) when an external magnetic field is applied. The values of $\Delta S_M$ increase with the magnitude of the external magnetic field. Higher magnitudes of the magnetic field $(d h)$ lead to greater increases in the peak value of $\Delta S$, thereby contributing more to the ability to cool the system.\\
Fig.\ref{Figure_9} shows the variation of the partial derivative of the magnetic field with respect to temperature and $\Delta S_{M}$ with respect to temperature. The number of Monte-Carlo steps used in this case is $10^{6}$. Our investigations are for the magnetic values of $0.1,0.2,0.3$ and $0.4$ in units of J. The different case study for different values of $\frac{J_y}{J_x}$ shows peaks at different temperatures. The figures are given in Fig. \ref{Figure_9}.
\section{Conclusion}
\label{S_8}
In this paper, motivated by the Wolff algorithm's speed over the Metropolis algorithm, we study the Ising model in both isotropic and anisotropic cases. We examine the behavior of different statistical quantities at various ($\frac{J_y}{J_x}$) values and present a study of the exact transition temperature variation using the Wolff algorithm through the Binder cumulant method. The studies reveal that with a decrease in the strength of the coupling constants, the transition temperature of the system decreases, while with an increase in the coupling constants, the transition temperatures increase, showing an almost linear relationship. Additionally, we review the hysteresis loop study for different anisotropic cases. It is found that at a particular temperature and for a constant number of Monte Carlo steps the width of the loops increases with the increase in the $\frac{J_y}{J_x}$ values. Furthermore, we demonstrate the response of different thermodynamic and magnetic quantities to varying fields. The behavior of the statistical quantities indicates that with an increase in the strength of the external magnetic field, the phase transition smears out, with the peaks of susceptibility and specific heat diminishing. We also investigate the magneto-caloric effect of the model. In the study of the magnetocaloric effect, the peak of the $\frac{\partial M}{\partial T} $ is shifting to higher temperatures as the value of the ratio $\frac{J_y}{J_x}$ is increasing because the transition temperature shifts to higher temperatures. In addition to that as the value of the external field is increasing the negative peak is decreasing and shifting to higher temperatures at the same time. In the study of $\Delta S_{M}$, it is found that a more pronounced change in the magnetic field will contribute more to the change in the magnetic entropy of the system. Thus, will give more contribution to the cooling power of the system.\\

\section*{Acknowledgments} The authors would like to thank for technical and software support of high performance computing (HPC) lab, VIT-AP University for providing computational resources.

\end{document}